\documentclass{article}

\usepackage{graphicx}  
\usepackage{amsmath}   
\usepackage{amssymb}   
\usepackage{bm}

\usepackage{relsize}  

\def\eq#1{{Eq.~(\ref{#1})}}

\def\frab#1#2{\left(\frac{#1}{#2}\right)}

\def\zpl{{zero-point-length}}

\def\bk#1#2#3{{\langle #1|#2|#3\rangle}}  

  \title{World-line Path integral for the Propagator expressed as an ordinary integral: Concept and Applications}
    
    \author{
  T. Padmanabhan\\
  {\small IUCAA, Pune University Campus,
  Ganeshkhind, Pune - 411 007, India.}\\
  {\textit {email: paddy@iucaa.in}}
  }
   
  \date{ }  

\begin{document}
  
  \maketitle
  
  \begin{abstract}
 The (Feynman) propagator $G(x_2,x_1)$ encodes the entire dynamics of a massive, free scalar field propagating in an arbitrary curved spacetime. The usual procedures for computing the propagator --- either as a time ordered correlator or from a partition function defined through a path integral --- requires  introduction of a field $\phi(x)$ and its action functional $A[\phi(x)]$. An alternative, more geometrical, procedure is to define a propagator in terms of the world-line path integral which only uses curves, $x^i(s)$, defined on the manifold. I show how the world-line path integral can be reinterpreted as an \textit{ordinary} integral by introducing the concept of effective number of quantum paths of a given length. Several manipulations of the world-line path integral becomes algebraically tractable in this approach.  In particular I derive an \textit{explicit} expression for the propagator $G_{\rm QG}(x_2,x_1)$, which incorporates the quantum structure of spacetime through a zero-point-length, in terms of the standard propagator $G_{\rm std}(x_2,x_1)$, in an arbitrary curved spacetime.  This approach also helps to clarify the interplay between the path integral amplitude and the path integral measure in determining the form of the propagator.  This is illustrated with several explicit examples. 
 \end{abstract}
 
 \section{Relativistic propagator from a geometrical perspective}
 
 Consider a free scalar field of mass $m$ which is propagating in a spacetime with metric $g_{ik}$ and is treated within the context of  quantum field theory in curved spacetime. I take the (generally accepted) point-of-view that the dynamics of such a field is completely contained in the standard relativistic (Feynman) propagator $G_{std}(x,y;m^2)$ or, equivalently, in the rescaled\footnote{As we shall see, there is some algebraic advantage in using $\mathcal{G}_{\rm std}$ rather than $G_{std}$. Of course, both contain the same amount of information in the case of a massive field, which I will be focusing on; the massless case can be treated by a limiting procedure and I will comment on it when relevant.} propagator $\mathcal{G}_{\rm std}(x,y;m)\equiv
mG_{std}(x,y;m^2)$. So if we have a prescription for computing $\mathcal{G}_{\rm std}(x,y;m)$, we can completely determine the dynamics of the field.

The usual procedure to determine $\mathcal{G}_{\rm std}(x,y;m)$ is to start with a Lagrangian for the scalar field, quantize the field and obtain the propagator from it. This itself can be done in two different ways. (i) One can identify the canonical momentum $\pi$ for $\phi$ and impose equal-time-commutation rules (ETCR) between them. If $\phi$ is expanded in terms  a set of mode functions, the ETCR will lead to the identification of  creation/annihilation operators. One can then construct a vacuum state, Fock basis etc.\footnote{In a time dependent background, there is the usual ambiguity of choice of positive/negative frequency mode functions, inequivalent vacuua etc. These are not relevant to the main thrust of the current discussion. Any one choice of mode functions and vacuum state is good enough for my purpose.} The propagator is then identified as the time-ordered  vacuum expectation value
$\bk{0}{T[\phi(x_2)\phi(x_1)]}{0}$. (ii) Alternatively one can find the  partition function $Z[J]$ by evaluating a path integral over $\phi$ of $\exp iA[\phi,J]$ after adding a source term $J(x)\phi(x)$ to the Lagrangian. The propagator can then be obtained as the second functional derivative of of $Z[J]$ evaluated at $J=0$.

Both these, standard, procedures use a field $\phi(x)$ and its action functional $A[\phi]$ as tools to arrive at the propagator $\mathcal{G}_{\rm std}$, which, ultimately, encodes  the entire dynamics. It is therefore useful to inquire whether we can determine  $\mathcal{G}_{\rm std}$ directly (and geometrically) without using the crutch of a field or its action functional. One motivation for this inquiry is the following: A purely geometrical definition of the propagator may be robust enough to survive (and be useful) at  scales close to --- but somewhat larger than --- Planck scales. (I call this regime mesoscopic; I will say more about it later on). 

Such a geometric approach is indeed possible because we know the differential equation and boundary conditions which $\mathcal{G}_{\rm std}$ satisfies. The relevant solution can be expressed in terms of the zero-mass-Schwinger-kernel (ZMSK) of the spacetime in the Schwinger's propertime representation as follows:\footnote{I will work in the Euclidean space(time) for mathematical convenience and will assume that the results in the pseudo-Riemannian spacetime arise through analytic continuation. This is particularly useful in the path integral representation discussed below.} 
 \begin{equation}
\mathcal{G}_{\rm std}(x,y;m)\equiv
mG_{std}(x,y;m^2)=\int_0^\infty m\ ds\ e^{-m^2s}K_{std}(x,y;s)
\label{a14}
 \end{equation} 
 where $K_{\rm std}$ is the standard ZMSK. This can be specified as a solution to a differential equation or as $K_{\rm std} (x,y;s) \equiv \bk{x}{e^{s\Box_g}}{y}$ where  $\Box_g$ is the Laplacian in the background space(time).
This  kernel is a purely geometric object, entirely determined by the background  geometry; the information about the scalar field is contained in the single parameter $m$. The $K_{\rm std} (x,y;s)$ has the structure (in $D=4$):
\begin{equation}
 K_{std}(x,y;s)\propto\frac{e^{-\bar\sigma^2(x,y)/4s}}{s^2}\left[1+ \text{curvature corrections}\right]
\end{equation} 
where $\bar\sigma^2(x,y)$ is the geodesic distance between the two events. The curvature corrections, encoded in the Schwinger-Dewitt expansion, will involve powers of $(s/L_{curv}^2)$. 

An equivalent,  more intuitive but  formal, definition of $\mathcal{G}_{\rm std}(x_1,x_2;m)$ is through a world-line path integral for a relativistic particle:
\begin{equation}
\mathcal{G}_{\rm std}(x_1,x_2;m) = \sum_{\rm paths\ \sigma}^{\mathcal{M}} \exp [-m\sigma (x_1,x_2)]
  \label{three}
 \end{equation}
 where $\sigma(x_1,x_2)$ is the geometrical length of a path connecting the two points $x_1, x_2$ and the sum is over all paths connecting these two events. 
 This is appealing because it uses the lengths of paths in space(time), which are purely geometrical entities, to give meaning to the propagator.
 Of course, the result of the path integral sum depends \textit{both} on on the summand $e^{-m\sigma}$ \textit{and} on the measure chosen for the path integral, indicated by $\mathcal{M}$ on top of the summation symbol in \eq{three}.  In flat spacetime, there is a straightforward  procedure to define $\mathcal{M}$: This is to define the sum over paths on a lattice and compute it --- with a suitable measure --- in the limit of zero lattice spacing \cite{tpqft,pid}. The lattice measure can then be chosen so that the sum will lead to the same propagator as in \eq{a14}. But, as is evident,
 defining and manipulating the measure, in a general curved background, is a nontrivial task. I stress that, the definition of the propagator through the path integral in \eq{three} is valid in arbitrary curved spacetime. In fact it is usually taken to be the definition of the propagator, in the world-line approach, and is often used with a quadratic action and a gauge function to ensure reparametrisation invariance.
 
I will now describe how this path integral can be converted into an \textit{ordinary} integral with a suitable integration measure which can be interpreted as the \textit{effective} number density of paths in space(time). This conceptual advance --- which will provide a completely geometrical and  useful description of the propagator --- is one of the key results of this paper. 
As we will see, this definition, for example, will  allow us to explore how the propagator gets modified when the quantum nature of the spacetime geometry is taken into account.

  To convert the path integral to an \textit{ordinary} integral, I will 
 introduce a Dirac delta function into the path integral sum in \eq{three} and use the fact that both  $\ell$ and $\sigma$  are positive definite, to obtain:\footnote{This works best in Euclidean sector because path lengths $d\sigma=\sqrt{g_{ab}dx^adx^b}$ are real. In an earlier work I have tried to do it with Lorentzian signature but it leads to ambiguities.}
 \begin{eqnarray}
\mathcal{G}_{\rm std}(x_1,x_2;m) &=& \int_{0}^\infty d\ell\ e^{-m\ell} \sum_{\rm paths\ \sigma}^{\mathcal{M}}\delta_D \left(\ell - \sigma (x_2,x_1)\right)\nonumber\\
 &\equiv& \int_{0}^\infty d\ell\ e^{-m\ell} N_{std}(\ell; x_2,x_1)
  \label{k2}
 \end{eqnarray}
 where we have defined the function $N_{std}(\ell; x_2,x_1)$ to be:
 \begin{equation}
N_{std}(\ell; x_2,x_1) \equiv \sum_{\rm paths\ \sigma}^{\mathcal{M}}\delta_D \left(\ell - \sigma (x_2,x_1)\right)
  \label{k3}
 \end{equation}
 The last equality in \eq{k2}
 describes the path integral as an \textit{ordinary} integral with an integration  measure $N_{std}(\ell;x_2,x_1)$.  This measure --- according to \eq{k3} --- can be thought of as counting the \textit{effective} number of paths\footnote{Of course, the \textit{actual} number of geometrical paths, of a given length connecting any two points in the Euclidean space, is either zero or infinity. But the \textit{effective} number of paths $N_{std}(\ell;x_2,x_1)$, formally defined as the inverse Laplace transform of $\mathcal{G}_{std}(x_2,x_1;m)$ (see \eq{k2}), will be a finite quantity.} of length $\ell$ connecting  $x_1$ and $x_2$. 
 This is a purely geometrical quantity defined in the space(time). 
 I will hereafter just write $N_{std}(\ell)$ for $N_{std}(\ell; x_2,x_1)$ etc., without explicitly displaying the dependence on the spacetime coordinates, for notational simplicity. The propagator $\mathcal{G}_{\rm std}(m)$ is just the Laplace transform of $N_{std}(\ell)$ from the variable $\ell$ to $m$;
 the measure $N_{std}(\ell)$ is  the inverse Laplace transform of $\mathcal{G}_{\rm std}(m)$ from the variable $m$ to $\ell$.
 
 To conclude this section, let me illustrate the explicit form of $N_{\rm free}(\ell)$ in the case of a free field in flat space.\footnote{\textit{Notation:} I will use the subscript `std' for functions pertaining to a classical gravitational background, \textit{not necessarily} a flat spacetime; for corresponding expressions evaluated  in the \textit{flat} spacetime, I will use the subscript `free'. In the later discussion the subscript `QG' will give the corresponding functions with quantum gravitational corrections. }   
 Translation invariance demands that both $\mathcal{G}_{\rm free}(x_1,x_2;m)$ and $N_{free}(\ell; x_2,x_1)$ will only depend on the difference $x\equiv (x_2-x_1)$. Fourier transforming the last equality in \eq{k2} with respect to $x$ we obtain a similar relation between $\mathcal{G}_{\rm free}(p;m)$ and $N_{free}(p,\ell)$ in the momentum space. The $N_{\rm free}(p,\ell)$ in the momentum space (in any dimension $D$) is given by the very simple expression:\footnote{From our definition, it follows that $N(x,\ell)$ has the dimensions of $L^{-D}$ in a $D$ dimensional space(time) which is the same as that of space(time) number density. Its Fourier transform $N(p,\ell)$ is dimensionless in all $D$.}
 \begin{equation}
 N_{\rm free}(p, \ell)= \cos(p\ell). 
 \label{simple}
 \end{equation}
Direct computation of this result will require lattice regularization of the sum in \eq{k3}. However, since the sum in \eq{three} can indeed be computed by lattice regularization \cite{tpqft,pid}, we can also compute  $N_{\rm free}(\ell)$ by an additional Laplace transform to arrive at \eq{simple}. 
 The result can be verified by taking the inverse Laplace transform of $\mathcal{G}_{\rm free}(p,m)=m(p^2+m^2)^{-1}$ or, more directly, by observing that:
 \begin{equation}
  \int_0^\infty d\ell\ e^{-m \ell} \cos p\ell =\frac{m}{m^2+p^2}= m G_{\rm free}(p^2,m^2)  =\mathcal{G}_{\rm free}(p^2,m) 
   \label{thirteen}
  \end{equation} 
Given $N_{\rm free}(p, \ell)$ in the momentum space,  $N_{\rm free}(\ell, x)$ in real space can be computed by evaluating the Fourier transform of $\cos p\ell$. The result is again quite simple.
  In $D=4$, we fins that (see Appendix \ref {app:path-measure} for calculational details):
  \begin{equation}
     N_{\rm free}(\ell,x) 
     =\frac{3}{4\pi^2}\  \, \frac{\ell \Theta[ \ell^2 - x^2]}{\left( \ell^2 - x^2\right)^{5/2}}
   \label{12mar}  
  \end{equation} 
  where $\Theta$ is the Heaviside theta function. 
It is amusing to note that $N_{\rm free}(\ell,x)$ vanishes for paths with length $\ell<x$ so that no such path will contribute to the Euclidean  integration measure. This is reminiscent of the fact that there are no geometrical paths with $\ell<x$ connecting the two events in the Euclidean space.

\section{Corrections to the propagator due to the quantum structure of spacetime}

 I have now defined the propagator entirely in terms of a geometric object $N_{std}(\ell; x_2,x_1)$ which could be thought of as the density of  effective number of paths between $x_2$ and $x_1$. As an application of this formalism, let us consider the following context.
 
 The description based on \eq{a14} --- which describes the  quantum field theory in a \textit{classical} background  spacetime --- is expected to breakdown when we probe the spacetime at length scales $\lambda\lesssim L_P$ where $L_P \equiv (G\hbar/c^3)^{1/2}$ is the Planck length. I will call this regime \textit{microscopic} and the regime of QFT in curved spacetime (CST), with $\lambda \gg L_P$), \textit{macroscopic}. 
 We need the full formalism of QG to study microscopic scales while QFT in CST is adequate for macroscopic scales.
 
 I am interested in the intermediate, \textit{mesoscopic} scales,   at which one can describe the spacetime in the usual continuum language and incorporate the prominent effects of QG by modifying\footnote{I want to work with a descriptor of the field dynamics  which is robust enough to survive (and be useful) at mesoscopic scales.  The propagator, described in terms of $N(\ell)$,  is a good choice for such a description.}
 the propagator $\mathcal{G}_{\rm std}$ to a quantum gravity corrected propagator $\mathcal{G}_{\rm QG}$. Such a description is expected to be valid  at length scales $\lambda \gtrsim C L_P$, with $C=10^3$, say, for definiteness. A factor of $10^3$ could allow for the continuum description to emerge, but --- at the same time --- be sensitive to the microscopic physics through a non-zero $L_P$. One cannot calculate  $G_{\rm QG}$  from first principles, without the full theory of quantum gravity. In the absence of such a luxury, I will use the  following working hypothesis to go forward. 
 
 It may be possible to capture the key effects of quantum gravity by introducing a \zpl\ to the spacetime \cite{ pid, zplimp,zplextra}. This is based on the idea that the \textit{dominant} effect of quantum gravity at \textit{mesoscopic} scales can be described by replacing the (squared) path length $\sigma^2(x_2,x_1)$  by $\sigma^2(x_2,x_1) \to \sigma^2(x_2,x_1)+ L^2$ where  $L^2$ is of the order of Planck area $L_P^2$. Such an idea is decades old and has been  explored extensively in the previous literature \cite{pid, zplimp, zplextra}. 
 All the same,  let me stress some aspects of this approach for the sake of conceptual completeness. This will be useful to readers  who are not sufficiently familiar with the earlier work   on this approach.

\begin{enumerate}

\item   The idea of bringing in the \zpl\ by the modification $\sigma^2 (x_2,x_1) \to \sigma^2(x_2,x_1) + L^2$ should be thought of as a working \textit{hypothesis} which is postulated to make progress, in the absence of of a complete theory of quantum gravity. This postulate is assumed to hold in an arbitrary curved spacetime.  idea has been introduced and explored extensively  in the  literature (for some early work, see \cite{pid, zplimp, zplextra, zpluse}; for more recent work, see \cite{zplrecent}) and I am describing some further consequences of this approach in this paper. 
 
 In principle, one should be able to derive this form of the propagator from a more complete theory of quantum gravity. For example, it can be obtained from the string theory \cite{fromstring} in a specific approximation; but for the purpose of this work, it is enough to consider it as a  postulate.

  \item By working directly with the propagator, we  bypass several nuances of standard QFT which may all require some  revision at mesoscopic scales. However, we know that both the dynamics \textit{and the symmetries} of a free quantum field, propagating in a curved geometry, is encoded in the Feynman propagator. So, if we understand how QG effects modify the propagator, we  can obtain a \textit{direct} handle on both the dynamics and the symmetries of the theory at mesoscopic scales. This is the motivation for  working directly with the propagator containing QG corrections, without worrying about  the (unknown) modifications to the standard formalism of QFT at mesoscopic scales.
  
  \item  As an example of the economy  gained by  this approach, let me stress the notion of diffeomorphism invariance in a curved geometry and --- as a special case --- the Lorentz symmetry in flat spacetime. The prescription $\sigma^2 (x_2,x_1) \to \sigma^2(x_2,x_1) + L^2$ is generally covariant because $L$ is  a, constant, (scalar) number.  In flat spacetime, our ansatz  will replace $(x_2 - x_1)^2$ by $(x_2 - x_1)^2 + L^2$ which is clearly Lorentz invariant. 
  The mere introduction of a constant length scale into the propagator will \textit{not} violate Lorentz invariance; this  should be obvious from the fact that the propagator for the massive scalar field does depend on the length scale $m^{-1}$ and is still perfectly Lorentz invariant. The results of detailed computations (see for example, the extensive set of computations in  \cite{zpluse}) explicitly demonstrate the  general covariance and Lorentz invariance of the procedure.
  
  This result is somewhat similar to that in, for example, LQG  which also contains a length scale but does not violate Lorentz invariance. Moreover, in our approach, the general covariance (and Lorentz invariance) \textit{is manifestly apparent} in the prescription $\sigma^2 (x_2,x_1) \to \sigma^2(x_2,x_1) + L^2$; so no special demonstration of this fact is required unlike, in the case of, for example, LQG \cite{carlos}.
  Some other prescriptions in the literature, for introducing a `minimal length', do create issues with Lorentz invariance but \textit{our prescription is (manifestly) generally covariant}. 
  
   \item   The action for the relativistic particle possesses a simple --- though not well-appreciated feature --- which plays a key role  in  this approach. The action, a priori,  is expected to be a functional of the form $A[x^a(\tau); x_1,x_2]$; that is, it is the \textit{functional} of the trajectory $x^a(\tau)$ and a \textit{function} of the end points $x_2$ and $x_1$. But, it can be expressed purely as a \textit{function} $A = A(\ell)$ of the length of the path $\ell[x^a(\tau); x_2,x_1]$, which carries the functional dependence on $x^a(\tau)$. This \textit{geometrical} structure of relativistic action is rather special; in fact, the usual action for the \textit{non}-relativistic particle \textit{cannot} be expressed  as a function of the length of the path. 
   
   As we shall see, it is this property which will allow us to translate the modification of path lengths in spacetime (by the addition of the \zpl) to the modification of the relativistic action. This, in turn, helps us to maintain all the relevant symmetries of the theory and directly compute the corrections to the propagator at mesoscopic scale, using the world-line path integral (bypassing e.g., the standard canonical quantization etc).

 \end{enumerate}

 After this rather long aside, let me come back to the main theme. It is easy to see how the introduction of \zpl\ into the  geometry changes the form of path integral \eq{three} for the propagator. The existence of the \zpl\ changes each path length $\sigma$ appearing in the amplitude to $(\sigma^2 + L^2)^{1/2}$. The quantum corrected propagator $\mathcal{G}_{QG}$ will then be given by the path integral sum:
 \begin{equation}
\mathcal{G}_{\rm QG}(x_1,x_2;m) = \sum_{\rm paths\ \sigma}^{\mathcal{M}} \exp [-m\sqrt{\sigma^2+L^2}]
  \label{threeQG}
 \end{equation}
 with the same measure $\mathcal{M}$. Once again, introducing the Dirac delta function and carrying out the steps which led to \eq{k2}, we will get:
 \begin{equation}
  \mathcal{G}_{\rm QG} (x_1,x_2; m) = \int_0^\infty d\ell \ N_{std}(\ell; x_1,x_2) \exp\left( - m \sqrt{\ell^2+L^2}\right)
  \label{b61}
 \end{equation} 
 where $N_{std}(\ell; x_1,x_2)$ is again defined through \eq{k3} and counts the effective number of paths. 
 This is the expression for the propagator in an (effective) quantum geometry with a \zpl.
 The  modification $\ell \to (\ell^2 + L^2)^{1/2}$ ensures that all path lengths are bounded from below by the \zpl\ as expected.\footnote{What happens to the classical relativistic particle if the action is modified from $m\ell$ to another function $A[\ell]$ monotonic in $\ell$? Since $\delta A=A'[\ell]\delta\ell$, the equations of motion does not change. This implies that, at least in the classical case, the dispersion relation $\omega^2=\bm{p}^2+m^2$ does not change by the addition of \zpl. In fact, it turns out that the dispersion relations for the excitations does not change  even when the propagator $\mathcal{G}_{QG}$ is obtained from a QFT but the discussion of this feature goes beyond the scope of this paper \cite{wip}.}
 
 Before proceeding further, let me again illustrate this result in the case of flat spacetime. Since  $N_{std}(\ell; x_1,x_2)$ depends only on $x=x_2-x_1$, so does $\mathcal{G}_{\rm QG} (x_1,x_2; m)=\mathcal{G}_{\rm QG} (x_2-x_1; m)$. Fourier transforming both sides of \eq{b61} we get a similar relation in the momentum space.  Now using the result that, in flat spacetime,  $N_{free}(\ell,p)=\cos p\ell$ in the momentum space, we get:
 \begin{equation}
  \mathcal{G}_{\rm QG}(p^2) =  \int_0^\infty d\ell \ e^{-m \sqrt{L^2+\ell^2}} \ \cos(p\ell) = \frac{mL}{\sqrt{p^2+ m^2}} K_1[ L\sqrt{p^2+m^2}]
   \label{fifteen1}
  \end{equation} 
  where $K_1(z)$ is the Bessel function of second kind and we have used a standard cosine transform (see p.16(26) of \cite{tit}). 
As to be expected, the $L\to 0$ limit leads to the standard expression  $\mathcal{G}_{\rm std}(p^2)=m( p^2+ m^2)^{-1}$ when we use the fact that in this limit $K_1(z)\to 1/z$. The result in real space is obtained by a Fourier transform  using standard integrals and we get:
\begin{equation}
 G_{\rm QG} (x) = 
 =\frac{1}{4\pi^2}\frac{m}{\sqrt{x^2+L^2}}K_1[m\sqrt{x^2+L^2}]=G_{\rm std} (\sqrt{x^2+L^2})
  \label{X10}
 \end{equation} 
The first equality comes from explicit Fourier transform of the result in \eq{fifteen1}. The second equality, however, tells us that the result  could have been ``guessed''. Since $G_{std}(x)$ in flat spacetime depends \textit{only} on $x^2$, we could have obtained $G_{QG}(x)$ by simply replacing $x^2$ by $x^2+L^2.$ As we shall see later, such simplicity does not occur in an arbitrary curved spacetime.

Similar results hold for the zero-mass-Schwinger-kernel (ZMSK). One can rewrite the result in \eq{fifteen1} in a different form, using another standard integral representation of the $K_1(z)$ function (see 3.324 (1) of \cite{gr}). We can write:
\begin{eqnarray}
   \mathcal{G}_{\rm QG}(p^2) &=& \frac{mL}{\sqrt{p^2+m^2}} \ K_1[L\sqrt{p^2+m^2}]
   =\int_0^\infty ds \ m\ \exp\left[-s(p^2+m^2) - \frac{L^2}{4s}\right] \nonumber\\
   &=&\int_0^\infty ds \ m\ e^{-m^2s}[K_{free}(s;p)e^{-L^2/4s}]
   \label{flatktok}
  \end{eqnarray}
First equality is just \eq{fifteen1}, the second equality uses an integral representation of the $K_1(z)$ function and in the third equality   
  we have used the flat space expression $K_{free}(s;p)=\exp(-sp^2)$ for the  ZMSK in the momentum space. This suggests that, at least in this case of flat space(time), the introduction of \zpl\ modifies the ZMSK by the replacement $K_{std}\to K_{QG}= K_{std}e^{-L^2/4s}$. Again, in flat spacetime we could have ``guessed'' this result; the ZMSK in flat spacetime,  $K_{free}(s;x)$, depends only on $x^2$  through a factor $\exp(-x^2/4s)$; so replacing $x^2$ by $x^2+L^2$ gives $K_{QG}(s;x)=K_{std}e^{-L^2/4s}$.
  Very surprisingly, as we shall see soon,  this result $K_{std}\to K_{QG}= K_{std}e^{-L^2/4s}$ holds even in arbitrary curved spacetime, even when  $K_{std}$ is not just a function of $x^2$. That fact cannot be ``guessed''.
  
 Let us now get back to the path integral expression in \eq{threeQG} for  $\mathcal{G}_{\rm QG} (x_1,x_2; m)$, which is valid in an \textit{arbitrary, curved} spacetime.
 It is a nontrivial problem to define a proper measure and do this summation in a curved spacetime even to obtain $\mathcal{G}_{\rm std} (x_1,x_2; m)$, let alone obtain
 $\mathcal{G}_{\rm QG} (x_1,x_2; m)$. Given the two path integrals:
 \begin{equation}
 \mathcal{G}_{\rm std} = \sum_{\rm paths\ \sigma}^{\mathcal{M}} \exp [-m\sigma];\qquad
 \mathcal{G}_{\rm QG} = \sum_{\rm paths\ \sigma}^{\mathcal{M}} \exp [-m\sqrt{\sigma^2+L^2}]
 \label{gandg}
 \end{equation}
 in a curved spacetime, the best we can hope is a relation between the two. Even this is very difficult to achieve in terms of path integrals. Here is where the introduction of $N(\ell)$ comes to our rescue.  We can now rewrite the propagators in \eq{gandg} in terms of integrals with measure $N(\ell)$
 instead of as path integrals. This leads to \eq{k2} for $\mathcal{G}_{\rm std} (x_1,x_2; m)$ and \eq{b61} for $\mathcal{G}_{\rm QG} (x_1,x_2; m)$:
 \begin{eqnarray}
   \mathcal{G}_{\rm std} (x_1,x_2; m) &=& \int_0^\infty d\ell \ N_{std}(\ell; x_1,x_2) \exp\left( - m \ell\right); \nonumber\\
   \mathcal{G}_{\rm QG} (x_1,x_2; m) &=& \int_0^\infty d\ell \ N_{std}(\ell; x_1,x_2) \exp\left( - m \sqrt{\ell^2+L^2}\right)
 \end{eqnarray} 
 If we eliminate the measure $N_{std}(\ell; x_1,x_2)$ between these two equations, we can express $\mathcal{G}_{\rm QG}$ directly in terms of $\mathcal{G}_{\rm std}$.
 
 While we do not know the form of $ N_{std}(\ell; x_1,x_2)$ in an arbitrary curved spacetime, we will often know  --- in practical contexts ---  the form of standard propagator $\mathcal{G}_{std}(x_1,x_2)$ [or the standard ZMSK $K_{std}(s;x_1,x_2)$] in a given spacetime. So, to actually compute the QG effects at mesoscopic scales, in a given curved spacetime, it will be useful if we can (a) express $\mathcal{G}_{QG}(x_1,x_2)$ in terms of $\mathcal{G}_{std}(x_1,x_2)$ and (b) express the quantum corrected ZMSK $K_{QG}(x_1,x_2; s)$ in terms of the standard ZMSK $K_{std}(x_1,x_2; s)$. Remarkably enough this can be done, without us knowing the explicit form of $N_{std}(\ell; x_1,x_2)$. 
 
 \textit{This is a key practical result of our approach,} which allows explicit computations without worrying about the measure for path integral etc. I will first quote the result, then describe some elementary consequences. The proof is given in the Appendix \ref{app:gtog}.
 
 The modification of the Schwinger kernel, due to introduction of \zpl, in arbitrary curved spacetime, is extremely simple: the QG effects involve the replacement:
 \begin{equation}
  K_{std}(s;x_1,x_2)\to K_{QG}=K_{std}(s;x_1,x_2)e^{-L^2/4s}
  \label{ktok}
 \end{equation} 
 so that \eq{a14} gets replaced by:
 \begin{equation}
  \mathcal{G}_{QG}(x,y;m)=\int_0^\infty m\ ds\ e^{-m^2s-L^2/4s}K_{std}(s; x,y)
  \label{a15}
 \end{equation} 
 Recalling that  the leading order behaviour of the ZMSK is $K_{\rm std} \sim s^{-2}\exp[-\bar\sigma^2(x,y)/4s]$ (where $\bar\sigma^2$ is the geodesic distance between the two events) we see that the modification in \eq{a15} amounts to the replacement $\bar\sigma^2 \to \bar\sigma^2 +L^2$ in the exponential factor. We have already verified in \eq{flatktok} that this is indeed the case in flat spacetime.
 
\textit{This result, valid in arbitrary curved spacetime,  is highly non-trivial and could not be ``guessed''.} 
 This is because, in a general spacetime  $K_{std}(s; x,y)$ will have  complicated dependences on $x,y$ and $K_{std}(s; x,y)$ \textit{cannot} be expressed as function of $\bar\sigma^2(x,y)$ alone; that is $K_{std}(s; x,y)\neq K_{std}(s; \bar\sigma^2(x,y))$. (Such a simplification arises   only in highly symmetric spacetimes like e.g., flat spacetime.) All the same, the replacement $\bar\sigma^2 \to \bar\sigma^2 +L^2$ in the \textit{leading exponential} correctly captures \textit{all} the effects of the \zpl.
 
 The relationship between $\mathcal{G}_{QG}$ and $\mathcal{G}_{std}$ is in the form of a convolution over the mass parameter. One can show that:
 \begin{equation}
  \mathcal{G}_{\rm QG} (x_2,x_1; m) = \int_0^\infty dm_0 \ \mathcal{P} \left[ m_0; m, L\right]\mathcal{G}_{\rm std} (x_2,x_1; m_0)
  \label{gtog}
 \end{equation} 
 where \footnote{One can carry out the differentiation in this expression and obtain two terms, one involving a Dirac delta function $\delta(m_0-m)$ and the other involving a product  $\Theta ( m_0-m)J_1[ L (m_0^2 - m^2)^{1/2}]$. It turns out to be often more convenient \textit{not} to carry out this differentiation and work with the expression in \eq{14aug1}.} 
 \begin{equation}
  \mathcal{P}[m_0; m,L] = - \frac{\partial}{\partial m} \left\{ \Theta ( m_0-m) J_0\left[ L \sqrt{m_0^2 - m^2}\right]\right\}
  \label{14aug1}
 \end{equation} 
This result is more useful in practical computations than \eq{ktok} because one more often knows the form of $\mathcal{G}_{\rm std} (x_2,x_1; m_0)$ than the form of $K_{\rm std} (x_2,x_1; s)$. 

In particular, 
if the background spacetime has certain symmetries, they constrain the structure of both $G_{\rm std}$ and  $G_{\rm QG}$ in a similar manner. For example, consider a class of homogeneous spacetimes (like e.g, the Friedmann universe) in which  $G_{\rm std} (x_1,x_2; m^2)=
 G_{\rm std} (\mathbf{x}_1-\mathbf{x}_2, t_1,t_2; m^2) $. It follows from \eq{gtog} that we will also have $G_{\rm QG} (x_1,x_2; m^2)=
 G_{\rm QG} (\mathbf{x}_1-\mathbf{x}_2, t_1,t_2; m^2) $. Therefore, one can Fourier transform both propagators with respect to spatial coordinates and obtain a  relation identical to \eq{gtog} in the momentum space:
 \begin{equation}
  \mathcal{G}_{\rm QG} (\bm{p},t_1,t_2; m) = \int_0^\infty dm_0 \ \mathcal{P} \left[ m_0; m, L\right]\mathcal{G}_{\rm std} (\bm{p},t_1,t_2; m_0)
 \end{equation} 
  Similarly, if the background spacetime is static, then one can Fourier transform both $\mathcal{G}_{\rm QG} $ and $\mathcal{G}_{\rm std}$ with respect to the difference in the time coordinates and obtain
  \begin{equation}
  \mathcal{G}_{\rm QG} (\omega,\bm{x}_1,\bm{x}_2; m) = \int_0^\infty dm_0 \ \mathcal{P} \left[ m_0; m, L\right]\mathcal{G}_{\rm std} (\omega,\bm{x}_1,\bm{x}_2; m_0)
 \end{equation} 
More complicated symmetries of $\mathcal{G}_{\rm std}$ can be handled in a similar manner. For example, in any maximally symmetric spacetimes, both $\mathcal{G}_{\rm std}$ and $\mathcal{G}_{\rm QG}$ will (essentially) depend on the geodesic distance and one can often deal with the Fourier transform with respect to the geodesic distance. 
This opens up several further avenues for concrete computation of QG effects in curved spacetime. I hope to explore these in a future work.

These expression in \eq{gtog} is  given as a convolution over the mass parameter in the propagator. It is also possible obtain other forms of relations from this result which could lead to algebraic simplifications or better intuitive understanding. I mention two alternative ways of relating $G_{QG}=\mathcal{G}_{QG}/m$ with $G_{std}=\mathcal{G}_{std}/m$; the proofs are in the Appendix \ref{app:gtog}. 

(1) By changing the integration variable in \eq{gtog} to $\mu\equiv (m_0^2-m^2)^{1/2}$ we can show that: 
\begin{equation}
 G_{\rm QG}(x_2,x_1; m) = - \frac{\partial}{\partial m^2} \int_0^\infty 2\mu\, d\mu\ J_0(L\mu) G_{\rm std} (x_2,x_1; \mu^2 +m^2)
 \label{gtognew}
\end{equation}

(2) It is also provide a higher dimensional interpretation of this this relation. Consider a fictitious $N=D+2$, Euclidean \textit{curved} space(time) with the metric
  \begin{equation}
  dS_N^2 = \left(g_{ab} (x) dx^a dx^b\right)_D + \delta_{AB}\ dX^A dX^B  \qquad \qquad (A,B = 1,2)
   \label{y10}
  \end{equation} 
  where we have added two ``flat'' directions, $X^A$ with $A=1,2$. Let $G^N_{std}(x, \bm{L}; y,\bm{0})$ be the standard propagator in the $N$ dimensional space with propagation in the fictitious directions being from origin to a point $\bm{L}$ with $|\bm{L}|$ being the \zpl. Then, one can show that the 
  propagator $G_{\rm QG}^D(x,y)$ in $D$ dimensions with \zpl\ $L$ is related to $G^N_{std}(x, \bm{L}; y,\bm{0})$(both for same mass $m$ which is not explicitly displayed) by:
 \begin{equation}
   G_{\rm QG}^D(x,y) = - 4\pi \frac{\partial}{\partial m^2} G^N_{std}(x,\bm{L}; y, \bm{0}) \bigg|_{\mathbf{L}^2=L^2}
    \label{y40}
  \end{equation} 
  The $N=D+2$ dimensional propagator, $G^N_{std}(x,\bm{L}; y, \bm{0})$, of course has a standard QFT interpretation in the curved spacetime.
 The \zpl\ in $D$ dimensions arises as the magnitude of the (fictitious) propagation distance in the extra dimensions.
  
 (3) One can also convert these relations into differential equations and show that: 
   \begin{equation}
  (-\Box^N + m^2)^{2} G^D_{QG} =4\pi \delta(x,y)\, \delta(\bm{L}) 
   \label{y80}
  \end{equation}
  and 
  \begin{equation}
   (-\Box^N + m^2) G^D_{QG} = 4\pi G^N_{std}(x, \bm{L}; y,\bm{0})
   \label{y90}
  \end{equation}
 These relations could form a basis for alternative interpretations of  $G_{QG}^D$. (Note that, in \eq{y80}, the $G^D$  refers to the propagator in $D$-dimensions while the Laplacian $\square^N$ is a $N=D+2$ dimensional one. The result is valid in arbitrary curved spacetime. In flat spacetime. 
 if you take the Fourier transform of \eq{y90}, the left hand side will lead to the square of the conventional momentum space propagator but the integration over the extra dimensions will lead to the correct result. The derivation of \eq{y80} in arbitrary curved spacetime and the explicit computation, in terms of the square of the  momentum space propagator in flat spacetime, are given in Appendix \ref{app:gtog}.)
 
 I conclude this section with two technical comments.
 
 The  procedure we have used, viz. the introduction of the effective 
number of paths and its QG generalization,  certainly works for the test scalar field propagating in an arbitrary curved space(time). For interacting fields (e.g a scalar field with $\lambda\phi^4$ coupling) it is not easy to represent the propagator in terms of a world-line path integral; therefore, the generalization of the current idea to interacting field theories is not straight forward. One possibly could capture some of the leading corrections to the \textit{perturbation theory} by using the $G_{QG}$ in the standard Feynman rules but it will miss non-perturbative effects. To tackle this problem, we first need to represent the interacting field theories, in curved spacetime, entirely in terms of the world-line approach --- which is a nontrivial task. But as long as we are only interested in probing the mesoscopic structure of spacetime using a simple quantum field, which \textit{is} our primary goal here, this approach is adequate.

 Let me conclude this section with a technical comment on the analytic continuation. Our approach starts with a curved space of Euclidean signature and obtains the results for the Lorentzian (pseudo-Riemannian) curved spacetime by an analytic continuation. It is is well-known that there is no rigorous, unique, correspondence between the set of all Euclidean metrics and the Lorentzian (pseudo-Riemannian) metrics. So, our procedure should be thought of as a \textit{general} prescription and ambiguous cases need to be handled on a case-by-case basis. In this context, the following two points need to be kept in mind: (a) This problem with analytic continuation arises whenever one uses Euclidean methods in curved spacetime and is \textit{not} specific to the discussion in this work.  (b) As long as one is concerned with the short-distance behaviour at mesoscopic scales, this problem can be circumvented. To make this notion precise, consider a non-singular spacetime with curvatures nowhere close to Planck values. That is, we assume, $L_P\ll L_{cur}$ where the curvature length scale defined by, say,  $L_{cur}^{-4}=R_{abcd}R^{abcd}$. Around any event $\mathcal{P}$, one can then introduce a locally inertial frame and use flat spacetime notion of analytic continuation at scales $L_P\lesssim x\ll L_{cur}$ which includes the mesoscopic scales we are interested in. This provides a way around the problem of analytic continuation in the regime we are interested in but, of course, it is not a solution to the broader issue of correspondence between Euclidean and pseudo-Riemannian spaces.

\section{Additional comments on the path measure}

  It should be obvious from the above discussion that manipulating $N_{\rm std}(\ell, x)$ is far easier than working with path integral measures and limiting processes. This has allowed us to obtain rather general results in the previous section in a concrete and simple manner. I will now discuss some further possible applications of this approach. 
  
  The standard propagator $\mathcal{G}_{\rm std}(x_1, x_2,m)$ was obtained by a path integral sum in \eq{three} with a specific measure $\mathcal{M}$ indicated as a superscript on the summation symbol. The same result translates to the ordinary integral in \eq{k2} in terms of the integration measure $N_{\rm std}(\ell,x_1, x_2)$. The definition of $N_{\rm std}(\ell,x_1, x_2)$ in \eq{k3} shows that it is defined using the path integral measure $\mathcal{M}$ and thus maintains a one-to-one correspondence with the choice of path integral measure. If we change the path integral measure, the functional form of $N_{\rm std}(\ell,x_1, x_2)$ will change and vice-versa. But for calculational purposes it is easy to change the form of the integration measure $N_{\rm std}(\ell, x_1, x_2)$ rather than the more abstractly defined path integral measure $\mathcal{M}$, with the implicit understanding that different choices of $N_{\rm std}(\ell, x_1, x_2)$ corresponds to different choices of the path integral measure $\mathcal{M}$. 
  
  This algebraic fact becomes significant when we realize that, in any physical situation, we are \textit{only} concerned with the propagator and not individually on the form of the integration measure $N(\ell, x_1, x_2)$ (or $\mathcal{M}$) and the form of the action used in the amplitude $\exp[-A(\ell)]$.  That is, in the expressions:
  \begin{equation}
\mathcal{G}(x_1,x_2;m) = \sum_{\rm paths\ \sigma}^{\mathcal{M}} \exp -A(\sigma)
=\int_{0}^\infty d\ell\ N(\ell; x_2,x_1) e^{-A(\ell)} 
\label{three29aug}
 \end{equation} 
 what matters for physics is the propagator  $\mathcal{G}(x_1,x_2;m)$ in the left-hand-side. It depends on both the form of $A(\sigma)$ as well as the measure $\mathcal{M}$ in the path integral or, equivalently, on the form of $A(\ell)$ and the path density $N(\ell; x_2,x_1)$ in the ordinary integral. The pair $\{\mathcal{M},A(\sigma)\}$
 or the pair $\{N(\ell),A(\ell)\}$, determines $\mathcal{G}(x_1,x_2;m)$ and \textit{different pairs can lead to the same propagator.} This fact is difficult to visualize or manipulate in terms of $\mathcal{M}$ but completely straightforward when we use $N(\ell)$
  
 Consider, as an example, the expressions for $\mathcal{G}_{\rm std}$ and $\mathcal{G}_{\rm QG}$ in \eq{k2} and \eq{b61}. In proceeding from  \eq{k2} to \eq{b61}, we \textit{postulated} that QG effects modify the world-line action by the replacement $m\ell \to m \sqrt{\ell^2+L^2}$. We then evaluated the sum over paths with the \textit{same original measure} $\mathcal{M}$, which is equivalent to using the original integration\footnote{Of course if you change $\ell$ to some other function $f(\ell)$ both in the measure and in the amplitude, you will change nothing in a definite integral.} measure $N_{\rm std}(x_1, x_2,\ell)$ in \eq{b61}. But, as I said, the physics only depends on the form of $\mathcal{G}_{\rm QG}$ and not individually on the form of integration measure or the action. For example, we could  have obtained the same result (i.e., the same $\mathcal{G}_{\rm QG} $) by keeping the path integral amplitude 
 to be the same (i.e., keeping the amplitude as $\exp -[m\sigma(x,x')]$) but introducing all the quantum gravity corrections on the path measure, by replacing $N_{std}$ by another measure $N_{QG}$. These two measures are related by the condition that we should get the same $\mathcal{G}_{\rm QG}(m)$. This requires:
 \begin{equation}
 \mathcal{G}_{\rm QG}(x_1,x_2;m)=\int_{0}^\infty d\ell\  N_{std}(\ell; x_2,x_1)e^{-m\sqrt{\ell^2+L^2}}
 =\int_{0}^\infty d\ell\  N_{QG}(\ell; x_2,x_1)e^{-m\ell}
  \label{k2new}
 \end{equation} 
 It is easy to determine  $N_{\rm QG} (\ell)$ by changing  the integration variable in the first integral in \eq{k2new} from $\ell$ to $\mu$  through $\mu^2 = \ell^2 + L^2$ and rewrite the integral as:
  \begin{equation}
  \mathcal{G}_{\rm QG} (x_1,x_2; m)= \int_0^\infty \frac{\mu \, d\mu}{\sqrt{\mu^2-L^2}} \, \Theta(\mu - L) \, N_{\rm std} (x_1,x_2; \ell = \sqrt{\mu^2 - L^2} ) \, e^{-m\mu}
   \label{twelve}
  \end{equation} 
 This is the form of the second integral \eq{k2new} in which we keep the standard form of the amplitude $\exp(-m\ell)$ but introduce the \zpl\ in the  path measure.
  The quantum corrected path measure $N_{\rm QG}(\mu)$ is then related to the standard  path measure $N_{\rm std}(\ell)$ by the simple relation 
  \begin{equation}
   N_{\rm QG}(x_1,x_2; \mu) = \frac{\mu}{\sqrt{\mu^2 - L^2}}\, N_{\rm std}\left[x_1,x_2; \ell = \sqrt{\mu^2-L^2}\right] \Theta(\mu - L)
   \label{nqgmu}
  \end{equation} 
The  $N_{\rm QG}(x_1,x_2; \mu)$, vanishes for paths with length less than the \zpl; they are treated as irrelevant to physics and does not contribute. I stress that (a) this  interpretation was possible only because $N_{\rm QG}$ etc. are  purely geometrical constructs, \textit{independent} of the mass $m$ of the field and (b) the result is valid in arbitrary curved spacetime, in which defining and manipulating path integral measures are difficult tasks. 
  Again let me quote the specific expression for   flat space(time) for illustration. Using the momentum space expressions  for $N_{\rm free}$ and $N_{\rm QG}$, and noting that $N_{\rm free}(p, \ell) = \cos(p\ell)$,  \eq{nqgmu} gives the corresponding $N_{\rm QG}(p,\ell)$ to be:
  \begin{equation}
  N_{\rm QG}(p, \ell) = \Theta(\ell - L) \frac{\ell \cos p\sqrt{\ell^2- L^2}}{\sqrt{\ell^2-L^2}}
   \label{fourteen}
  \end{equation}

  As a final application, let me consider a particular form of the QG modified action, which has been used in previous literature. This is given by: 
  \begin{equation}
   A(\sigma) = -m \left[\sigma + \frac{L_0^2}{\sigma}\right]
  \end{equation} 
  which has the nice property that it is invariant under $\sigma \to L_0^2/\sigma$ which was called `duality' in previous literature. It turns out that one can actually compute the path integral sum with this action (see \cite{pid}), using a \textit{non-standard} measure $\mathcal{M}'$ and arrive at the \textit{same} QG corrected propagator $\mathcal{G}_{\rm QG}$.  Our conversion of the path integral into an ordinary integral, in terms of an integration measure $N(\ell, x_2, x_1)$, allows us to see more transparently how this result comes about. By changing the variable from $\ell $ to $\bar \ell$ with $\ell = \bar \ell - (L_0^2/\bar \ell)$ --- so that the integration range $\ell = (0, \infty)$ can be mapped to $\bar \ell = (\ell_0, \infty)$ --- we can re-express the integral for $\mathcal{G}_{\rm QG}$ in the desired form. Elementary algebra gives 
  \begin{equation}
 \mathcal{G}_{\rm QG}(x_2, x_1,m) \equiv \int_0^\infty d\ell \ N_{\rm std} (\ell, x_2, x_1) \, e^{-m \sqrt{\ell^2 + L^2}} = \int_0^\infty d\sigma \ N_{\rm QG} (\sigma,x_2, x_1)\, e^{-m\left(\sigma + \frac{L^2_0}{\sigma}\right)}
  \end{equation} 
  with $L_0 = L/2$ and 
  \begin{equation}
  N_{\rm QG}(\sigma, x_2, x_1) = \Theta(\sigma - L_0) \left( 1+ \frac{L_0^2}{\sigma^2}\right) \, N_{\rm std}\left(\ell = \sigma - \frac{L_0^2}{\sigma}, x_2, x_1\right)
  \end{equation} 
  This result explicitly relates the two integration measures used in the two cases and remains valid in arbitrary curved spacetime. We see that $N_{\rm QG}(\sigma)$ restricts the paths to those with length $\sigma > \ell$. Such manipulations  are not so transparent when one works with the path integral measure $\mathcal{M}$ and $\mathcal{M}'$.

 \section{Summary of results}
 
 I summarize below the key new results and their significance. 
 \begin{itemize}
  \item The propagator for a spinless particle of mass $m$ in an arbitrary curved background with metric $g_{ab}(x)$ encodes the full quantum dynamics of the system. It can be represented as a purely geometric object in terms of a world-line path integral. The measure for such a path integral is difficult to define and manipulate in an arbitrary curved spacetime. I introduce the concept of effective number of paths $N(\ell; x_1,x_2)$ and re-write the world-line path integral as an ordinary integral:
  \begin{equation}
   \mathcal{G}(m; x_1,x_2) = \sum_{{\rm paths}\, \sigma}^{\mathcal{M}} e^{-A[\sigma]} = \int_0^\infty d\ell\ N(\ell; x_1,x_2) e^{-A[\ell]}
   \label{sum1}
  \end{equation} 
  where $A[\sigma] = m \sigma (x_1,x_2) $ is the standard action for the relativistic particle. As it turns out, this conversion of a path integral measure $\mathcal{M}$ to an ordinary integral measure   $N(\ell; x_1,x_2)$  is  a key technical and conceptual advance introduced in this paper. This idea works only  because the action for a relativistic particle is a geometrical entity and can be expressed entirely in terms of the path length $\sigma(x_1,x_2)$. 
  
  \item I illustrate this concept in terms of flat spacetime and show that the effective number of paths has an extremely simple expression in momentum space and is given by $N(\ell,p) = \cos p\ell$. 
  
  \item This approach really comes alive when one considers the propagator $\mathcal{G}_{\rm QG}(m; x_1,x_2)$ at mesoscopic scales, incorporating the effects of \zpl\ by the modification $\sigma \to \sqrt{\sigma^2 + L^2}$. This propagator $\mathcal{G}_{\rm QG}(m; x_1,x_2)$ is given by exactly the same expression as in \eq{sum1} with $A(\sigma) = m\sigma $ replaced by $A(\sigma) = m \sqrt{\sigma^2 + L^2}$. The path integral sum is now intractable but the ordinary integral in terms of $N(\ell; x_1,x_2)$ comes to our rescue. For example, one can explicitly compute the QG corrected propagator when the background metric is flat and obtain, in momentum space, the result:
  \begin{equation}
  \mathcal{G}_{\rm QG}(p^2) =  \int_0^\infty d\ell \ e^{-m \sqrt{L^2+\ell^2}} \ \cos(p\ell) = \frac{mL}{\sqrt{p^2+ m^2}} K_1[ L\sqrt{p^2+m^2}]
   \label{fifteen2}
  \end{equation} 
  It is non-trivially difficult to do this explicit computation, even in flat spacetime, using path integral measure. 
  
  \item The concept of effective number of paths turns out to be much more useful in curved spacetime. In an arbitrary curved spacetime we can express both the standard propagator and the QG corrected one in terms of $N(\ell; x_1,x_2)$  by converting the respective path integrals into ordinary integrals and thereby obtaining:
  \begin{eqnarray}
   \mathcal{G}_{\rm std} (x_1,x_2; m) &=& \int_0^\infty d\ell \ N_{std}(\ell; x_1,x_2) \exp\left( - m \ell\right); \nonumber\\
   \mathcal{G}_{\rm QG} (x_1,x_2; m) &=& \int_0^\infty d\ell \ N_{std}(\ell; x_1,x_2) \exp\left( - m \sqrt{\ell^2+L^2}\right)
 \end{eqnarray} 
   It is now possible to eliminate $N(\ell; x_1,x_2)$ between these two relations, by some algebraic gymnastics, and relate $ \mathcal{G}_{\rm QG} (x_1,x_2; m)$ directly to $\mathcal{G}_{\rm std} (x_1,x_2; m)$. These relations can be expressed in many different forms; for e.g., one can show that 
    \begin{equation}
 G_{\rm QG}(x_1,x_2;m^2) = - 4\pi \frac{\partial}{\partial m^2} \int\frac{d^2k}{(2\pi)^2}\, e^{i\bm{k\cdot L}}\, G_{\rm std} (x_1,x_2; k^2 +m^2)
 \end{equation} 
 where $\bm{L}$ is a 2-dimensional vector with magnitude equal to the \zpl. 
 
 \item One can also relate  $G_{\rm QG}^D(x,y)$ in $D$-dimensions with the standard propagator in a fictitious space of $N=D+2$ dimensions withe metric
 $ds^2=g_{ab}(x)dx^adx^b+\delta_{AB}dX^AdX^B$, where we have added two ``flat directions`` $X^A$ with $A=1,2$. We can prove that:
 \begin{equation}
   G_{\rm QG}^D(x,y) = - 4\pi \frac{\partial}{\partial m^2} G^N_{std}(x,\bm{L}; y, \bm{0}) \bigg|_{\mathbf{L}^2=L^2}
    \label{y400}
  \end{equation} 
The $N=D+2$ dimensional propagator, $G^N_{std}(x,\bm{L}; y, \bm{0})$, of course has a standard QFT interpretation in the curved spacetime.
 The \zpl\ in $D$ dimensions arises as the magnitude of the (fictitious) propagation distance in the extra dimensions.
  We also have the differential relations:
   \begin{equation}
  (-\Box^N + m^2)^{2} G^D_{QG} =4\pi \delta(x,y)\, \delta(\bm{L}) 
   \label{y800}
  \end{equation}
  and 
  \begin{equation}
   (-\Box^N + m^2) G^D_{QG} = 4\pi G^N_{std}(x, \bm{L}; y,\bm{0})
   \label{y900}
  \end{equation}
between the two propagators.

 \end{itemize}

\section*{Acknowledgment}

  My research  is partially supported by the J.C.Bose Fellowship of Department of Science and Technology, Government of India.

\appendix
 
 \section{Appendix: Calculational details}
 
 \subsection{Path measure in real space}\label{app:path-measure}

Given the  path measure in the momentum space, $N_{\rm free}(\ell, p) = \cos p\ell$, we can find the measure $N_{\rm free}(\ell, x)$ in real space by evaluating the D-dimensional  Fourier transform of $\cos p\ell$. To do this, we start with the standard result for the Fourier transform of spherically symmetric function. If 
  \begin{equation}
 F(k) = \int d^D\bm{x}\ f(|\bm{x}|) \, e^{-i\bm{k\cdot x}}
  \end{equation}
  then we can write
  \begin{equation}
k^{\frac{D-2}{2}} F(k) = (2\pi)^{D/2} \int_0^\infty J_{\frac{D-2}{2}} (kr) \, r^{\frac{D-2}{2}}\ f(r)\, r \, dr
\label{foursevena}
  \end{equation}
  This allows us to write the relevant Fourier transform as:
  \begin{equation}
  N_{free}(\ell,x) = \int_0^\infty \frac{d^Dp}{(2\pi)^D}\  e^{ip.x} \cos p\ell = \frac{1}{(2\pi)^{D/2}} \frac{1}{x^\alpha} \int_0^\infty dp \ p^{\alpha + 1} J_\alpha (p x) \cos \ell p
  \end{equation} 
  where $\alpha = (D/2) - 1$. We next evaluate this integral using the standard cosine transform  (see, for e.g., page 45, 1.12 (12) of \cite{tit})
  \begin{equation}
 \int_0^\infty dx\ (\cos xy)\, x^{\nu + 1} J_\nu (ax) = \frac{2^{\nu + 1} \sqrt{\pi}\, a^\nu y \Theta[y -a]}{\Gamma\left(-\frac{1}{2} - \nu\right) (y^2 - a^2)^{\nu + \frac{3}{2}}}
 \label{tit45}
  \end{equation} 
  This gives the result 
   \begin{equation}
  N_{free}(\ell,x) = \frac{1}{\pi^{\frac{D-1}{2}}} \ \frac{\Theta[\ell^2-x^2]}{\Gamma\left(-\frac{D-1}{2}\right) }\ \frac{\ell}{[\ell^2 - x^2]^{\frac{D+1}{2}}}
  \label{nreal}
  \end{equation} 
  which reduces to the expression quoted in the text when $D=4$. (Strictly speaking the integral in \eq{tit45} is defined only for $-1<{\rm Re}\nu<-{1/2}$ but can be analytically continues for other $\nu$, as often done in dimensional regularization.) 
  Note that $N_{\rm free}(\ell, x)$ vanishes for $\ell < x$; paths, with lengths less than the geometrical length  between the two points, do not contribute to the path integral which is rather nice feature. 
  
  One can directly verify that this expression leads to the correct massive propagator in real space which, of course, is obvious from the fact that $N_{\rm free}(\ell, x)$ is the Fourier transform of $N_{\rm free} (\ell,p)$ and we know that the latter gives the correct propagator in the momentum space.  To verify this directly  we need the integral
   \begin{equation}
 \int_1^\infty dt\ \frac{t \ e^{-zt}}{(t^2-1)^{\frac{1}{2} - \nu}} = \frac{\Gamma\left(\nu + \frac{1}{2}\right)}{\sqrt{\pi}}\, \frab{2}{z}^\nu \, K_{\nu+1}(z)
  \end{equation}
  which can be obtained from a standard result (see, 8.432 (3) of \cite{gr}) by differentiating with respect to $z$ and using the recursion relation for $K_\nu(z)$ (see page 929 (13) of \cite{gr}).  Using this integral and the expression for $N_{free}(\ell,x)$ in \eq{nreal}. we find that (with $\mu \equiv (1/2) (D+1)$)
  \begin{align}
   \int_0^\infty d\ell \ N_{\rm free}(\ell,x)\ e^{-m\ell}  
    &= \int_x^\infty d\ell \frac{1}{\pi^{\mu -1}} \, \frac{1}{\Gamma(1-\mu)} \, \frac{\ell\ e^{-m\ell}}{(\ell^2 - x^2)^\mu}\nonumber\\
    &= \frac{x^{2(1-\mu)}}{\pi^{\mu -1}\sqrt{\pi}} \, \frab{2}{mx}^{\frac{1}{2} - \mu} K_{\mu- \frac{3}{2}} (m x)  = \mathcal{G}_D(mx)
  \end{align} 
  which is indeed $m$ times the standard massive propagator $\mathcal{G}_D(mx) =mG_D(mx)$ in D-dimensions. In the case of $D=4$, this reduces to 
  \begin{equation}
  \mathcal{G}(x) = m \, G(x) = \frac{m}{4\pi^2} \frab{m}{x}\, K_1(mx)
  \end{equation} 
  which is the familiar expression.

 \subsection{Relation between $\mathcal{G}_{QG}$ and $\mathcal{G}_{std}$}\label{app:gtog}
 
I will now sketch the proof of \eq{gtog} and \eq{ktok} which involves slightly nontrivial manipulations of integrals over Bessel functions. I begin with two easily provable identities:
\begin{equation}
 \frac{e^{-m\sqrt{\ell^2 + L^2}}}{\sqrt{\ell^2 + L^2}} = \int_m^\infty dm_0\ e^{-m_0\ell} \, J_0\left[ L \sqrt{m_0^2 - m^2}\right] 
  \label{14aug2}
 \end{equation}
 and 
 \begin{equation}
 \frac{1}{2s} e^{-m^2s - \frac{L^2}{4s}} = \int_m^\infty dm_0\ m_0 e^{-m_0^2s} J_0 \left[ L \sqrt{m_0^2-m^2}\right]
  \label{14aug6}
 \end{equation}
The first identity in \eq{14aug2} can be proved by changing the integration variable on the right hand side to $x\equiv m_0/m$ and using a standard integral (see, 6.616 (2) of \cite{gr}) 
 \begin{equation}
 \int_1^\infty dx\ e^{-\alpha x} J_0\left[ \beta \sqrt{x^2-1}\right] = \frac{e^{-\sqrt{\alpha^2 + \beta^2}}}{\sqrt{\alpha^2 + \beta^2}}
  \label{14aug3}
 \end{equation} 
The second identity in \eq{14aug6} can again be proved by changing the integration variable on the right hand side to $x=m_0/m$ and using a result  derivable from 6.614 (1) of \cite{gr}:
 \begin{equation}
   \int_0^\infty 2k dk \ J_0(kL) e^{-sk^2} = \frac{1}{s} \exp\left( -\frac{L^2}{4s}\right)
   \label{ythirty}
  \end{equation}
Differentiating both sides of \eq{14aug2} with respect to $m$, we obtain 
 \begin{align} 
 e^{-m\sqrt{\ell^2 + L^2}}  &= -\frac{\partial}{\partial m} \int_m^\infty dm_0\ e^{-m_0 \ell}  J_0\left[L  \sqrt{m_0^2-m^2}\right] \nonumber\\
 &=  \int_0^\infty dm_0 \ e^{-m_0\ell}\, (-1) \frac{\partial}{\partial m} \left\{ \Theta(m_0-m) J_0\left[ L \sqrt{m_0^2 - m^2}\right]\right\}
  \label{14aug4}
 \end{align}
 This gives
 \begin{equation} 
 e^{-m\sqrt{\ell^2 + L^2}} = \int_0^\infty dm_0\ e^{-m_0\ell} \mathcal{P}[m_0; m, L]
  \label{14aug5}
 \end{equation}
 with $\mathcal{P}$ defined by \eq{14aug1}. Similarly, differentiating both sides of \eq{14aug6} with respect to $m$ and manipulating as before we get 
 \begin{equation}
 m e^{-m^2 s - \frac{L^2}{4s}} = \int_0^\infty dm_0 \ m_0 e^{-m_0^2 s } \mathcal{P} [m_0; m,L]
  \label{14aug7}
 \end{equation}
 Notice that the right hand sides of \eq{14aug7} and  \eq{14aug5} have very similar structures with $e^{-m_0\ell}$ replaced by $m_0e^{-m_0^2 s}$.  
 
 The results in  \eq{gtog} and \eq{a15}, which we need to prove, can now be obtained in a straightforward manner as follows:
  Multiply both sides of \eq{14aug5} by $N(\ell)$ and integrate over $\ell$  to get:
  \begin{align}
 \mathcal{G}_{\rm QG} &= \int_0^\infty d\ell \ N(\ell) \, e^{-m\sqrt{\ell^2 + L^2}} = \int_0^\infty dm_0\ \mathcal{P} [ m_0; m, L] \int_0^\infty d\ell \ N(\ell) \, e^{-m_0 \ell}\nonumber\\
 &= \int_0^\infty dm_0\ \mathcal{P} [ m_0; m, L]\mathcal{G}_{\rm std} (m_0)
  \label{14aug8}
 \end{align}
 where $\mathcal{P} [ m_0; m, L]$ is defined by \eq{14aug1}, reproduced here for ready reference:
 \begin{equation}
  \mathcal{P}[m_0; m,L] = - \frac{\partial}{\partial m} \left\{ \Theta ( m_0-m) J_0\left[ L \sqrt{m_0^2 - m^2}\right]\right\}
  \label{14aug1a}
 \end{equation} 
  This gives \eq{gtog}. To obtain \eq{a15}, we write $\mathcal{G}_{\rm std}$ in terms of $K_{\rm std}(s)$. This leads to 
 \begin{align}
  \mathcal{G}_{\rm QG} &= \int_0^\infty dm_0\ \mathcal{P} [ m_0; m, L]\int_0^\infty ds\ e^{-m_0^2s} K_{std}(s) m_0 \nonumber\\
  & = \int_0^\infty ds\  K_{std}(s)\int_0^\infty dm_0\ m_0 \,  e^{-m_0^2s}\mathcal{P} [ m_0; m, L]\nonumber\\
  & =  \int_0^\infty ds\ K_{std}(s) m e^{-m^2 s - \frac{L^2}{4s}}
  \label{14aug9}
 \end{align}
 In arriving at the last equality we have used the result in \eq{14aug7}. This proves \eq{a15}. 
 
 To obtain the relation between $G_{\rm QG} = \mathcal{G}_{\rm QG} / m$ and $G_{\rm std} = \mathcal{G}_{\rm std} / m$ we can proceed as follows. We pull  the derivative with respect to $m$ (arising from the expression in \eq{14aug1a}) out of the integral sign in \eq{14aug8} and change the variable of integration to $k$ with $k^2 \equiv m_0^2 - m^2$. This leads to the relation:
 \begin{equation}
 \mathcal{G}_{\rm QG} (m) = - \frac{\partial}{\partial m}  \int_0^\infty \frac{kdk}{\sqrt{k^2+m^2}}\ J_0(Lk)\ \mathcal{G}_{\rm std}(m^2+k^2)
 \end{equation} 
 where $\mathcal{G}_{\rm std}(m^2+k^2)$ is the standard propagator with $m^2$ replaced by $m^2+k^2$. Using the fact that $\mathcal{G}(M) = MG(M)$ for any mass parameter $M$, this is equivalent to 
 \begin{align}
  G_{\rm QG} (m) &= - \frac{1}{m}\frac{\partial}{\partial m} \int_0^\infty \frac{kdk}{\sqrt{k^2+m^2}}\ J_0(Lk)\ \sqrt{k^2+m^2} \, G_{\rm std}(m^2+k^2)\nonumber\\
   &= - \frac{\partial}{\partial m^2} \int_0^\infty 2k\, dk\ J_0(Lk)\ G_{\rm std} ( k^2 +m^2)
  \label{3sep27}
 \end{align}
 This is the relation quoted in the main text; see \eq{gtognew}. Further, for any function $f(k)$ which depends only on the magnitude of the 2-dimensional vector $\bm{k}$, we have the identity
 \begin{equation}
\int\frac{d^2k}{(2\pi)^2} \, e^{i\bm{k\cdot L}}\, f(k) = \frac{1}{2\pi}\int_0^\infty k\,dk\ J_0(kL)\,  f(k) 
 \end{equation} 
 Using this \eq{3sep27} can be expressed in the form 
 \begin{equation}
 G_{\rm QG}(m) = - 4\pi \frac{\partial}{\partial m^2} \int\frac{d^2k}{(2\pi)^2}\, e^{i\bm{k\cdot L}}\, G_{\rm std} ( k^2 +m^2)
 \end{equation} 
 where $\bm{L}$ is a 2-dimensional vector with magnitude equal to the \zpl. From this it is possible to obtain a higher dimensional interpretation of $G_{\rm QG}$. However, the result is probably more transparent when obtained from first principles and I will provide such a derivation:

 I will now work in $D$ dimensional space(time) and express $G_{\rm QG}^D$ in $D$-dimensions in terms of the \textit{standard} propagator $G_{\rm std}^N$ in $N=D+2$ dimensions. To do this, let us consider a fictitious $N=D+2$, Euclidean \textit{curved} space(time) with the metric
  \begin{equation}
  dS_N^2 = \left(g_{ab} dx^a dx^b\right)_D + \delta_{AB}\ dX^A dX^B  \qquad \qquad (A,B = 1,2)
   \label{y1}
  \end{equation} 
  where we have added two ``flat'' directions, $X^A$ with $A=1,2$. (The metric $g_{ab}$ in $D$ dimensions, of course depends only on the $D$ coordinates $x^a$.) The $N$-dimensional Schwinger  kernel for $m=0$ (ZMSK) now factorizes and we can write  
\begin{equation}
 K^N_{m=0} \equiv \bk{x,\bm{L}}{e^{s\Box_N}}{y,\bm{0}}
 =\left(\frac{1}{4\pi s}\right) e^{-L^2/4s} \, K^D_{m=0}(x,y; s) \, ;\qquad L^2 \equiv L_AL^A
   \label{y2}
  \end{equation} 
  Therefore, the corresponding $N$-dimensional, massive, Schwinger kernel becomes  
\begin{equation}
 K^N(x,\bm{L}; y,\bm{0}; s) = \left(\frac{1}{4\pi s}\right) e^{-(L^2/4s) - m^2s} \, K^D_{m=0}(x,y; s)
   \label{y3}
  \end{equation} 
  The corresponding $N$-dimensional massive propagator is obtained by integrating this kernel over $s$ in the range 0 to $\infty$. This is almost the same as $G_{\rm QG}$ but for the extra factor $(1/4\pi s)$.  This factor can be  taken care of by differentiating the propagator with respect to $m^2$. We then find that  
\begin{equation}
 -4\pi \frac{\partial}{\partial m^2} \, G_{\rm std}^N(x,\bm{L}; y,\bm{0}) = \int_0^\infty ds\ e^{-(L^2/4s) - m^2s} \, K^D_{m=0}(x,y; s)
  = G_{\rm QG}^D (x,y) 
  \end{equation}
  So, we have related the quantum corrected propagator $G_{\rm QG}^D(x,y)$ in $D$ dimensions to the standard Klein-Gordan propagator in the fictitious $N=D+2$ space with the metric in \eq{y1}, through the relation:
  \begin{equation}
   G_{\rm QG}^D(x,y) = - 4\pi \frac{\partial}{\partial m^2} G^N_{\rm std}(x,\bm{L}; y, \bm{0}) \bigg|_{\mathbf{L}^2=L^2}
    \label{y4}
  \end{equation} 
  The $N=D+2$ dimensional propagator, $G^N_{\rm std}(x,\bm{L}; y, \bm{0})$, of course has a standard QFT interpretation in the curved spacetime.
 The \zpl\ in $D$ dimensions arises as the magnitude of the (fictitious) propagation distance in the extra dimensions. This proves \eq{y40} in the main text.
  
  The derivative of standard Klein-Gordan propagator with respect to $m^2$ --- which appears in the right hand side of \eq{y4} ---can be related to the `transitivity integral' for the propagator $G^N_{\rm std}$. To see this, consider the following integral  (with the notation  $d^N\bar z \equiv d^Dz\, d^2Z$): 
  \begin{align}
  \int d^N\bar z \, G^N_{\rm std}(x,\bm{L}; z,\bm{Z}) \, G^N_{\rm std}(z,\bm{Z}; y,\bm{0}) &\nonumber\\
  &\hskip -10em = \int d^N\bar z \ \bk{x,\bm{L}}{(-\Box^N + m^2)^{-1}}{z,Z} \bk{z,Z}{(-\Box^N + m^2)^{-1}}{y,\bm{0}}\nonumber\\
  &\hskip -10em = \bk{x,\bm{L}}{(-\Box^N + m^2)^{-2}}{y,\bm{0}} 
   \label{y5}
  \end{align}
 We can, however, write:
   \begin{equation}
   \bk{x,\bm{L}}{(-\Box^N + m^2)^{-2}}{y,\bm{0}} = -  \frac{\partial}{\partial m^2} \, G^N_{\rm std}(x,\bm{L}; y,\bm{0})
   \label{y6}
  \end{equation}
 Combining this result with \eq{y4} we find that 
  \begin{equation}
  G^D_{QG} = (4\pi) \bk{x,\bm{L}}{(-\Box^N + m^2)^{-2}}{y,\bm{0}} \bigg|_{\mathbf{L}^2=L^2}
   \label{y7}
  \end{equation}
  This result, in turn, can expressed in the form of either of the following differential equations for $G^D_{QG}$, viz.: 
  \begin{equation}
  (-\Box^N + m^2)^{2} G^D_{QG} =4\pi \delta(x,y)\, \delta(\bm{L}) 
   \label{y8}
  \end{equation}
  and 
  \begin{equation}
   (-\Box^N + m^2) G^D_{QG} = 4\pi G^N_{\rm std}(x, \bm{L}; y,\bm{0})
   \label{y9}
  \end{equation}
 These are valid  any curved space(time). We can now write $G^N_{\rm std}(x, \bm{L}; y,\bm{0})$ as the standard vacuum expectation value of time ordered products of the KG field operators in $N=D+2$ space(time); then the QG corrected propagator in $D$ dimensional space is given by a solution to \eq{y9}. This proves \eq{y80} and \eq{y90}. 
  
 We can, of course,  verify that, in flat space(time) either of these equations lead to the correct $G_{\rm QG}$. For example, if we Fourier transform either \eq{y8} or \eq{y9}, we will find that $G^D_{\rm QG}$ can be expressed as the integral
  \begin{equation}
  G^D_{QG} (x, \bm{L}; 0,\bm{0}) = 4\pi \int\frac{d^Dk \, d^2K}{(2\pi)^N} \ \frac{e^{ikx}\, e^{i\bm{K\cdot L}}}{(k^2 + m^2 + K^2)^2}
   \label{y1000}
  \end{equation}
  The square appearing in the denominator can be taken care of by the usual trick of differentiating the expression with respect to $m^2$. Performing the 2-dimensional integral over the measure $d^2K = K dK d\theta$ we get the result in terms of the Bessel function $J_0(KL)$:
  \begin{align}
  G^D_{QG} (x, \bm{L}; 0,\bm{0}) &= -\frac{\partial}{\partial m^2} \int \frac{d^Dk}{(2\pi)^D} \int_0^\infty K dK\int_0^{2\pi}\frac{d\theta}{\pi} \,  \frac{e^{ikx}\, e^{i{K L\cos\theta}}}{k^2 + m^2 + K^2} \nonumber\\
  &= -\frac{\partial}{\partial m^2} \int \frac{d^Dk}{(2\pi)^D}\, e^{ikx}\int_0^\infty 2K dK\ \frac{J_0(KL)}{k^2 + m^2 + K^2} 
   \label{y11}
  \end{align}
  Therefore the $D$-dimensional Fourier transform  $G^D_{\rm QG} (k,\bm{L})$ of 
  $G^D_{\rm QG}(x, \bm{L}; 0,\bm{0})$  is given by 
  \begin{equation}
   G^D_{\rm QG} (k,\bm{L}) = - \frac{\partial}{\partial m^2} \int_0^\infty 2K dK \ \frac{J_0(KL)}{k^2 + m^2+K^2}
  \end{equation}
  To perform the integral over $K$ we write the denominator in the exponential form, leading to: 
  \begin{align}
 -\frac{\partial}{\partial m^2} \int_0^\infty 2K dK &\ J_0(KL) \int_0^\infty  ds\ e^{-s(k^2+m^2)} \, e^{-sK^2}\nonumber\\
 & = \int_0^\infty ds\ s e^{-s(k^2+m^2)} \times \ \int_0^\infty 2K \, dK\ J_0(KL) e^{-sK^2}
   \label{y12}
  \end{align}
 Finally, we use the  identity 
  \begin{equation}
   \int_0^\infty 2K dK \ J_0(KL) e^{-sK^2} = \frac{1}{s} \exp\left( -\frac{L^2}{4s}\right)
   \label{ythirty00}
  \end{equation} 
  to recover the standard result
  \begin{equation}
  G^D_{\rm QG} (k,{L}) = \int_0^\infty ds\ e^{-s(k^2+m^2) - (L^2/4s)}
   \label{y13}
  \end{equation}

 \end{document}